\documentclass{article}

\usepackage{PRIMEarxiv}

\usepackage[utf8]{inputenc} 
\usepackage[T1]{fontenc}    
\usepackage{hyperref}       
\usepackage{url}            
\usepackage{booktabs}       
\usepackage{amsfonts}       
\usepackage{nicefrac}       
\usepackage{microtype}      
\usepackage{lipsum}
\usepackage{fancyhdr}       
\usepackage{graphicx}       
\usepackage{multirow} 
\graphicspath{{media/}}     

\title{Enhancing Cryo-EM Density Map Segmentation in Phenix for Improved Atomic Model Building}

\author{
  Chenwei Zhang \\
  Department of Computer Science \\
  The University of British Columbia  \\
  Vancouver, Canada\\
  \texttt{zhang.chenwei@hotmail.com}
}

\begin{document}
\maketitle

\begin{abstract}
We introduce PhenixCraft, a fully automated pipeline for building atomic models from cryo-EM density maps. By integrating AlphaFold predictions, we enhance the map-segmentation step in Phenix during model building, addressing challenges posed by noise and artifacts that traditionally hinder this step. Our results demonstrate PhenixCraft's superior performance in TM-scores and sequence accuracy, significantly improving upon the limitations and inefficiencies of traditional model building using Phenix.
\end{abstract}


\section{Introduction}

The resolution revolution of cryogenic electron microscopy (cryo-EM) technique \cite{kuhlbrandt2014resolution,henderson2015overview,de2021cryo} opens a new era for automated atomic model building from cryo-EM density maps \cite{zhang2025review,diiorio2023novel,empot,phenix,modelangelo,deeptracer} to determine spatial structures of biological molecules. 
However, accurately constructing protein structures from their density maps is challenging owing to the large and complex protein structures with flexible regions,regions of low resolution within the maps, and the presence of artifacts and noise.  
The methodologies for constructing protein atomic models involve optimization-driven approaches such as Phenix \cite{phenix} and Rosetta \cite{rosetta} that employ comprehensive physics-based and statistical potential-based optimization algorithms in the volumes of cryo-EM density maps to identify the type and position of residues of the protein, and deep learning-oriented approaches such as ModelAngelo \cite{modelangelo} and DeepTracer \cite{deeptracer} that utilize neural networks to predict the corresponding residues of the volumes in cryo-EM density maps and subsequently assemble the final protein configuration. Among these methods, the Phenix software called \emph{phenix.map\_to\_model} has been widely studied and used as baselines for advanced model building methods, which has built atomic models for over 600 complexes, including proteins and nucleic acids. However, we noticed that the predicted protein structures by using Phenix often contain substantial residues with incorrect amino acid types, large portions of incomplete fragments, and incorrect chain assignments. 
This can be attributed to the deficiencies in the map-segmentation stage, where noise and artifacts within the map hinder the algorithm's ability to properly segment the map into distinct regions for subsequent model building.

To address this limitation, in this paper we introduce an alternative approach to boost the effectiveness of map segmentation by leveraging the predicted units (chains of the protein complex) from foundation models like AlphaFold2, which delivers highly accurate predictions of monomeric protein structures \cite{alphafold}.
Specifically, we employ AlphaFold-predicted structures to segment certain map regions rich in interpretative and informational content, while efficiently filtering out noisy components. Importantly, these AlphaFold-predicted structures are not included in the final atomic models. Additionally, we have developed a fully automated pipeline, named \emph{PhenixCraft}, for protein model building.

\section{Related work}

Phenix is a comprehensive software suite for determining macromolecular structures from cryo-EM data. Recent advances in high-resolution cryo-EM density maps have led to significant developments in model building, notably by Terwilliger et al. with the introduction of \emph{phenix.map\_to\_model}. This fully automated tool constructs 3D protein and nucleic acid structures from high-resolution maps, typically finer than 4.5 Ångströms ({\AA}), without manual intervention. The process includes several critical steps. 
\paragraph{(\romannumeral 1) Map sharpening.} Experimental cryo-EM density maps are initially sharpened using the \emph{phenix.auto\_map} tool, enhancing map detail and connectivity \cite{autosharpen}. This is achieved by optimizing the global sharpening factor \cite{sharpen} applied to Fourier coefficients of the map up to its effective resolution. A blurring factor is also introduced to reduce high-resolution noise, ensuring clarity and interpretability. More recently, deep learning–based approaches to map sharpening have demonstrated promising performance, offering data-driven alternatives to traditional frequency-domain methods \cite{emready,cryosamu,deepemhancer,struc2mapgan}. 

\paragraph{(\romannumeral 2) Map segmentation.}
This process begins by segmenting cryo-EM maps at densities below an automatically determined threshold. 
Connected regions exceeding this threshold are identified and grouped, often linked by symmetry. A unique set of density regions is selected from each group of symmetry-related entities for the subsequent model-building algorithms. This selection process is optimized to yield a structure that is compact and highly connected.
The threshold choice is critical, optimized based on the specific volume of the map above the threshold, desired region size, and available symmetry information. However, noise can complicate threshold determination, potentially leading to misaligned density regions, incorrect amino acid-type prediction, and atom localization.      
This segmentation algorithm may also falter if map symmetry is less significant, potentially yielding density regions corresponding to multiple molecular units instead of a single one. 

\paragraph{(\romannumeral 3) Model building and refinement.} After map segmentation, the selected density regions are input into a model-building algorithm to construct polypeptide backbones, identify secondary structures, and refine the model to its final structure. Each selected region is initially modeled as proteins or nucleic acids, employing various map-interpretation methods. The modeled regions are then combined; overlapping interpretations are removed, retaining only the best-fitting models. An assembled model representing the entire density map is created, applying reconstruction symmetry information. Finally, the constructed model is refined to enhance its accuracy and reliability.

\section{Methodology}

In this section we introduce the enhanced map-segmentation method and the PhenixCraft workflow to improve the performance of atomic model building using Phenix.

\subsection{Enhancement of map segmentation} 

To address the map-segmentation challenges discussed in the related work section, our approach begins by extracting density regions primarily associated with individual chains. 
We first dock one or several chain models into the map using a convolution-based shape search to locate map areas that resembles the model. This is accomplished with the \emph{phenix.dock\_in\_map} tool \cite{dockinmap}.
Subsequently, we construct a molecular mask with a radius of 3 {\AA} around the docked model using the \emph{phenix.map\_box} tool \cite{mapbox}. This mask helps to isolate the docked parts by excluding density volumes outside of the selected area. The resulting maps are characterized by high density and compactness, which facilitates the subsequent model-building steps. 
This enhanced map-segmentation method not only significantly reduces large noise volumes within the map, thereby improving the accuracy of amino acid-type prediction, but also guides the sequence assignment. This ensures that the resulting partial maps are precisely correlated with their corresponding chains, preventing inappropriate chain assignment.

\subsection{The PhenixCraft workflow}

\begin{figure}[ht]
\centering
\includegraphics[width=\linewidth, trim={0cm 3cm 0cm 0cm}, clip]{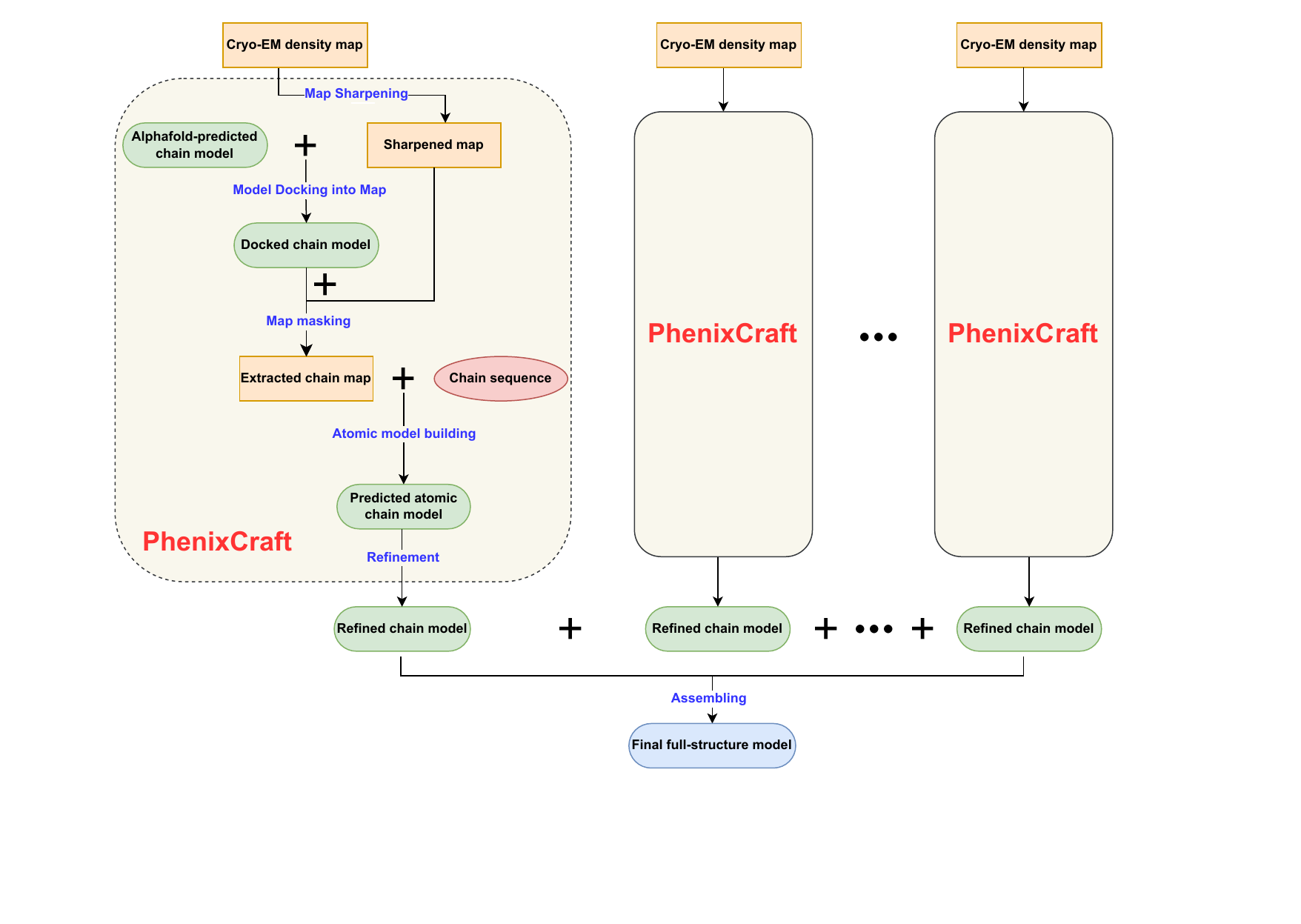}
\caption{The architecture of the PhenixCraft pipeline. All maps are shown in rectangular boxes with pale orange backgrounds. All models are shown in rounded rectangular boxes with pale green backgrounds except the final full-structure atomic model with a light blue background. Sequences are shown in the elliptic boxes with pale red backgrounds.}
\label{fig:phenixcraft}
\end{figure}

To enhance user accessibility, we have developed a fully automated pipeline that integrates our map-segmentation method with Phenix's model-building module, as shown in Figure \ref{fig:phenixcraft}.
This process begins with the sharpening \emph{phenix.auto\_sharpen} to maximize connectivity and detail. Subsequently, chain models predicted by AlphaFold are docked into the sharpened map to extract the corresponding chain maps.
Model building for each chain is then conducted employing \emph{phenix.map\_to\_model}, which takes the segmented chain model and its corresponding sequence as inputs. 
Once the initial chain model is constructed, it is refined to correct register errors by aligning the density at side-chain positions with sequence information, utilizing the \emph{phenix.insertions\_deletions} tool.
This process is repeated until all chain models are constructed. The full-structure atomic model is then assembled from these chain models.
In cases where the protein complex is a homomer, with all chains identical, we construct a single chain model and utilize symmetry information to assemble the final structure.

\subsection{Datasets} \label{dataset}
In our experiment, we constructed atomic models for 20 cryo-EM density maps at resolutions ranging from 1.8 {\AA} to 4.3 {\AA}, as listed in Supplementary Table \ref{tab:s1}. These maps correspond to a variety of protein structures: 5 monomers with a single chain, 9 homomers consisting of multiple identical chains, and 6 heteromers composed of various types of chains. The cryo-EM density maps were sourced from \texttt{Electron Microscopy Data Bank} \cite{EMDB} (EMDB). Their corresponding PDB structures, used as ground-truth references for benchmarking, were obtained from \texttt{RCSB Protein Data Bank} \cite{PDB} (RCSB PDB). AlphaFold-predicted chain models were sourced from \texttt{AlphaFold Protein Structure Database} \cite{alphafold}. For chain models not available in the database, we generated them locally using the latest version of AlphaFold. The atomic models constructed by Phenix were sourced from its website \cite{phenixdata}, and these structures were utilized for comparison with those generated by PhenixCraft. Proteins including 7TU5, 7UNL, and 8E2L, not available from the website, were produced using \emph{phenix.map\_to\_model}.

\subsection{Implementation} 
PhenixCraft was developed using Python 3.10. The Phenix software with version 1.20 was employed for map sharpening, model docking, map masking, and atomic model building and refinement. AlphaFold was installed and implemented locally in its latest version and can also be run on ColabFold \cite{colabfold}, a free and accessible platform for protein folding. Visualization of cryo-EM density maps and protein atomic models was performed using UCSF ChimeraX \cite{chimeraX}.
The TM-score \cite{tmscore} was calculated using the open-source library MM-align \cite{mmalign} to assess the structural similarity between the predicted structures and their references. Code and datasets will be released upon publication.

\section{Results}

\subsection{Quantitative comparison}

To quantitatively evaluate our method, we constructed 20 protein atomic models from our dataset (see section \ref{dataset}). We benchmarked the performance of PhenixCraft against \emph{phenix.map\_to\_model} based on two metrics: TM-score and sequence match percentage.
The results, presented in Table \ref{tab:comp}, demonstrate that PhenixCraft achieves an average TM-score of 0.324 and an average sequence match percentage of 31.12 across the 20 models, while significantly outperforming \emph{phenix.map\_to\_model}, suggesting that PhenixCraft represents a substantial improvement in atomic model building from cryo-EM density maps.

\begin{table}[ht]
\vspace{2em}
\centering
\caption{Comparison of PhenixCraft and \emph{phenix.map\_to\_model}.}
\begin{tabular}{|c|c|c|}
\hline
& TM-score & Seq. match \% \\
\hline
PhenixCraft & \textbf{0.324} & \textbf{31.12} \\
\hline
phenix.map\_to\_model & 0.186 & 24.56\\
\hline
\end{tabular}
\label{tab:comp}
\vspace{2em}
\end{table}

\begin{figure}[ht]
\centering
\includegraphics[width=\linewidth, trim={0cm 9cm 6cm 0cm}, clip]{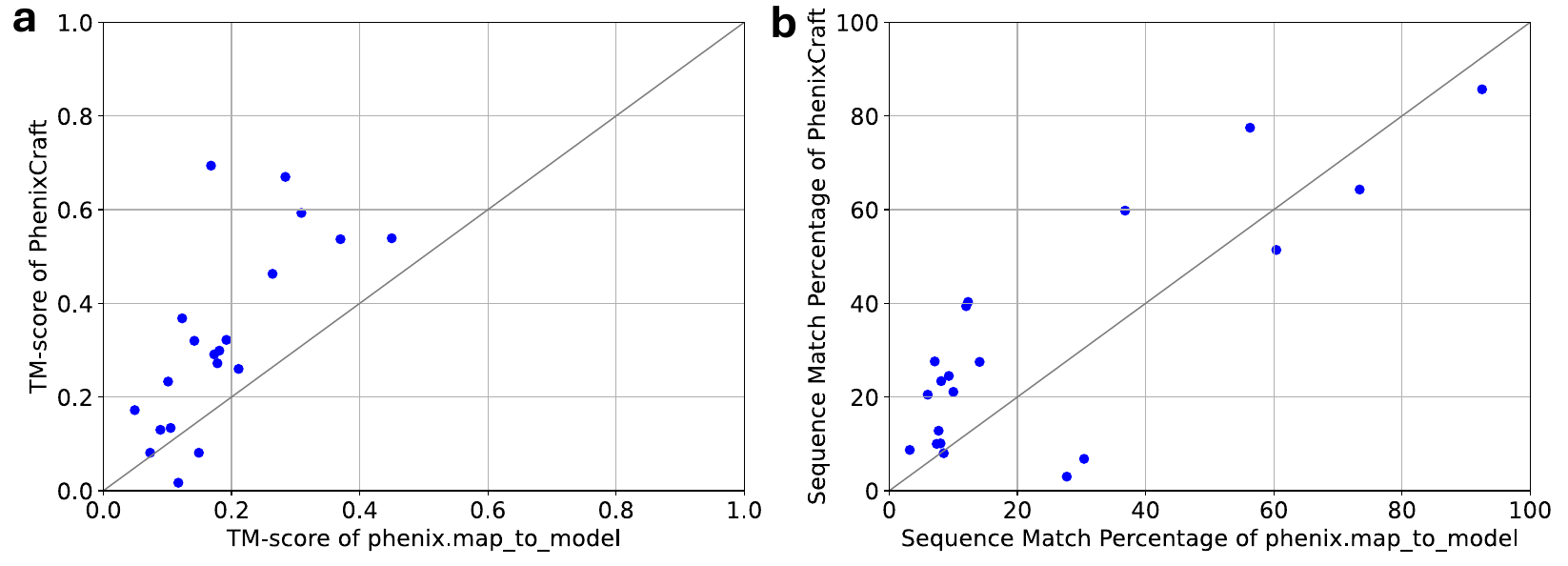}
\caption{Evaluation of the built models against the reference PDB structure for PhenixCraft and \emph{phenix.map\_to\_model}. Comparison of the (\textbf{a}) TM-score and (\textbf{b}) sequence match percentage.}
\label{fig:plot}
\end{figure}

Figure \ref{fig:plot} provides comparative analyses between PhenixCraft and \emph{phenix.map\_to\_model}. Panel a compares the TM-scores from \emph{phenix.map\_to\_model} (x-axis) to those from PhenixCraft (y-axis). Data points above the diagonal line (18 out of 20 points) indicate that PhenixCraft consistently achieves higher TM-scores, suggesting a more accurate model construction. Panel b examines the sequence match percentage for the same two methods. Data points predominantly above the diagonal line (14 out of 20 points) highlight that PhenixCraft often reports higher sequence matching percentages, demonstrating more precise predictions of amino acid types during model building.
The complete assessment scores are detailed in Supplementary Table \ref{tab:s1}.

\subsection{Visual comparison}

\begin{figure}[!ht]
\centering
\includegraphics[width=\linewidth, trim={0cm 1cm 19cm 0cm}, clip]{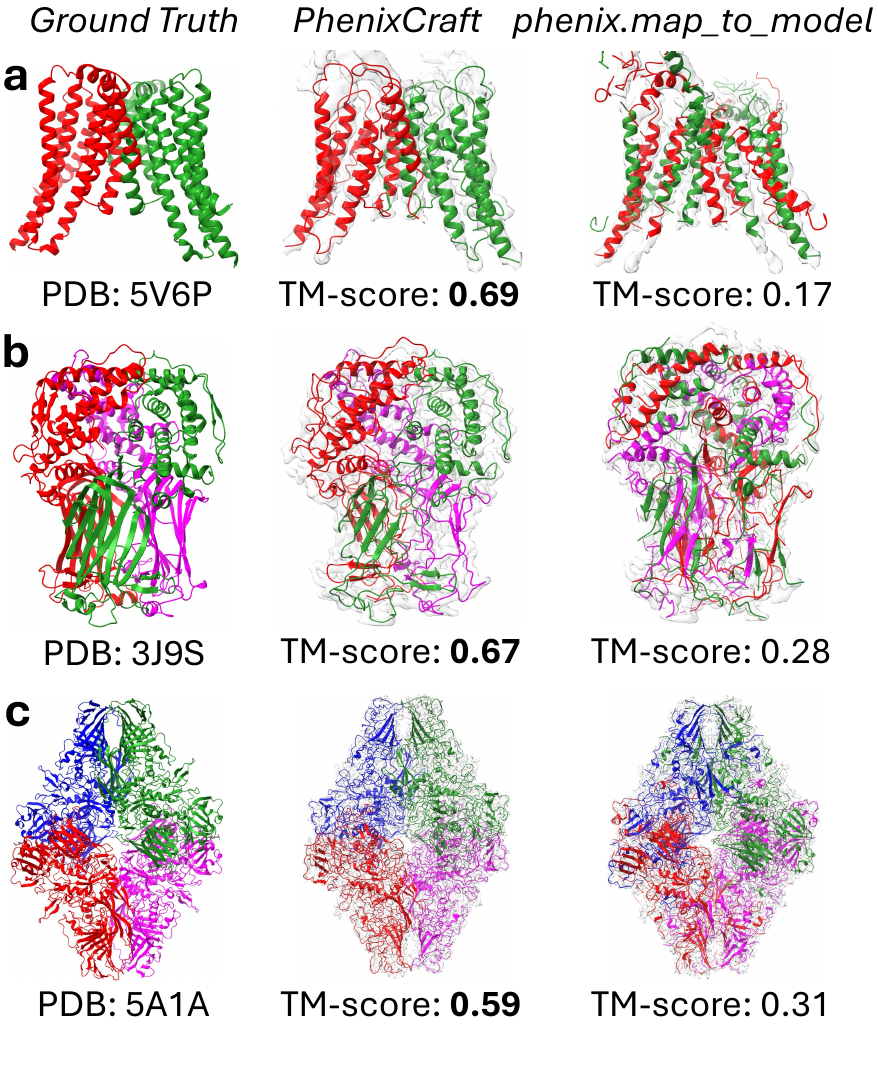}
\caption{
Constructed models by PhenixCraft and \emph{phenix.map\_to\_model}. Chains are colored separately. The TM-scores for each model are listed alongside. 
\textbf{a.} ERAD-associated E3 ubiquitin-protein ligase HRD1 (EMDB ID: 8637; PDB ID: 5V6P; reported resolution: 4.1 {\AA}).
\textbf{b.} Rotavirus VP6 (EMDB ID: 6272; PDB ID: 3J9S; reported resolution: 2.6 {\AA}).
\textbf{c.} Beta-galactosidase in complex with a cell-permeant inhibitor (EMDB ID: 2984; PDB ID: 5A1A; reported resolution: 2.2 {\AA}).
}
\label{fig:vis}
\end{figure}

We then visualized three representative atomic models constructed using PhenixCraft, as shown in Figure \ref{fig:vis}. We observed that PhenixCraft successfully built higher-quality models for EMDB entries 8637, 6272, and 2984 compared to \emph{phenix.map\_to\_model}. These models achieved TM-scores above 0.5, indicating a close match with the ground-truth models.
Although the models constructed by \emph{phenix.map\_to\_model} were complete, many amino acids were incorrectly assigned to chains where they do not belong, resulting in inaccurate models and very low TM-scores. In addition, some models built by \emph{phenix.map\_to\_model} were incomplete and their secondary structures poorly constructed, which also contributed to low TM-scores (see Supplementary Figure \ref{fig:a1}).

\clearpage

\section{Conclusion}

In this work, we present PhenixCraft, a novel pipeline for constructing atomic models from cryo-EM density maps, which integrates AlphaFold's predictions to enhance the map-segmentation step in Phenix. We extensively tested our approach across 20 various density maps, and our results demonstrate that PhenixCraft outperforms \emph{phenix.map\_to\_model} in terms of TM-score and sequence accuracy, highlighting its effectiveness in constructing models that closely resemble true protein structures. Future efforts will aim to refine this pipeline, particularly the docking step, where the chain models occasionally failed to dock correctly in the map.

\clearpage
\bibliographystyle{unsrt}  
\bibliography{references}

\appendix
\newpage
\section*{Supplementary Material}

\renewcommand{\thetable}{S\arabic{table}}
\renewcommand{\thefigure}{S\arabic{figure}}
\setcounter{table}{0} 
\setcounter{figure}{0}

\begin{figure}[ht]
\centering
\includegraphics[width=\linewidth, trim={0cm 4cm 8cm 0cm}, clip]{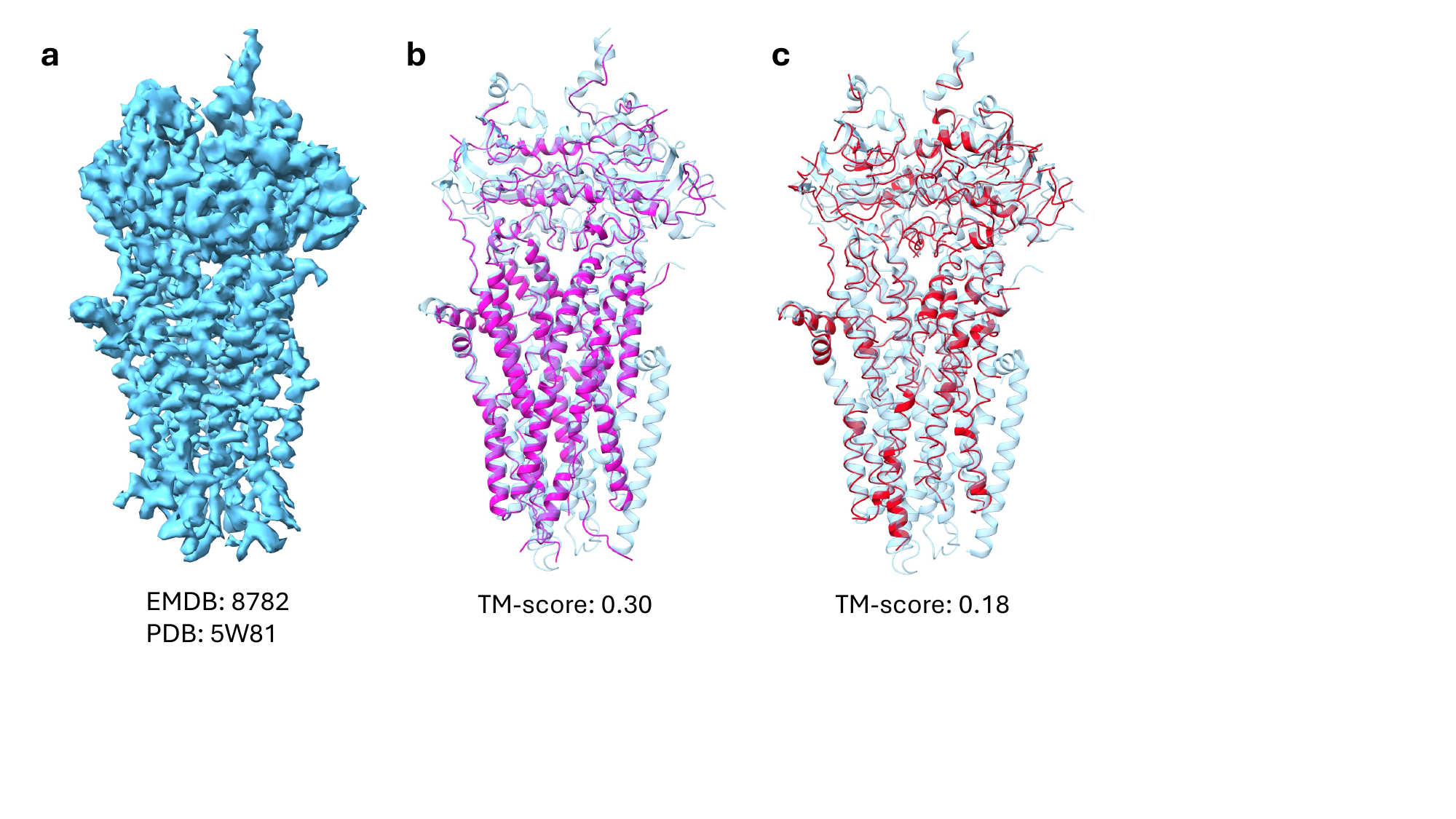}
\caption{
In (\textbf{a}), the experimental density map is colored blue.
Constructed models by PhenixCraft (\textbf{a}) and \textit{phenix.map\_to\_model} (\textbf{b}) are displayed, with the TM-scores listed alongside. The transparent blue structures in both (\textbf{a}) and (\textbf{b}) represent the ground-truth structure. This structure is the phosphorylated, ATP-bound structure of zebrafish cystic fibrosis transmembrane conductance regulator, with an EMDB ID: 8782 and a PDB ID: 5W81. The reported resolution is 3.37 \AA.}
\label{fig:a1}
\end{figure}

\newpage

\begin{table}[ht]
\centering
\caption{A list of the TM-scores and sequence match percentages by PhenixCraft and \textit{phenix.map\_to\_model} for 20 constructed atomic models from test cryo-EM density maps. Higher scores are highlighted in bold.}
\label{tab:s1}
\begin{tabular}{lcccccc}
\hline
\multirow{2}{*}{EMDB} & \multirow{2}{*}{PDB} & \multirow{2}{*}{Resolution (\AA)} & \multicolumn{2}{c}{TM-score} & \multicolumn{2}{c}{Seq. match \%} \\
\cline{4-7}
 &  &  & PhenixCraft & \textit{phenix.map\_to\_model} & PhenixCraft & \textit{phenix.map\_to\_model} \\
\hline
5K12 & 8194  & 1.8  & \textbf{0.537} & 0.370 & \textbf{24.5} & 9.3  \\
7TU5 & 26126 & 2.1  & \textbf{0.463} & 0.264 & \textbf{23.4} & 8.1  \\
5A1A & 2984  & 2.2  & \textbf{0.593} & 0.309 & \textbf{20.5} & 6.0  \\
7UNL & 26626 & 2.45 & \textbf{0.539} & 0.450 & 8.0  & \textbf{8.5}  \\
3J9S & 6272  & 2.6  & \textbf{0.670} & 0.284 & \textbf{10.1} & 8.0  \\
5UJA & 8560  & 3.34 & \textbf{0.322} & 0.192 & 6.8  & \textbf{30.4} \\
5W81 & 8782  & 3.37 & \textbf{0.299} & 0.181 & \textbf{8.7}  & 3.2  \\
5SZS & 8331  & 3.4  & \textbf{0.130} & 0.089 & 85.7 & \textbf{92.5} \\
7RZY & 24783 & 3.5  & \textbf{0.172} & 0.049 & 64.3 & \textbf{73.4} \\
8E2L & 27842 & 3.51 & \textbf{0.081} & 0.073 & \textbf{39.4} & 12.0 \\
5UAR & 8461  & 3.73 & \textbf{0.260} & 0.211 & \textbf{12.8} & 7.7  \\
5MZ6 & 3583  & 3.8  & \textbf{0.291} & 0.173 & \textbf{77.5} & 56.3 \\
5V7V & 8642  & 3.9  & \textbf{0.272} & 0.178 & \textbf{59.8} & 36.8 \\
5U1D & 8482  & 3.97 & \textbf{0.368} & 0.123 & 51.4 & \textbf{60.4} \\
5XSY & 6770  & 4.0  & \textbf{0.320} & 0.142 & 3.0  & \textbf{27.7} \\
5UZ7 & 8623  & 4.1  & \textbf{0.233} & 0.101 & \textbf{27.6} & 7.1  \\
5V6P & 8637  & 4.1  & \textbf{0.694} & 0.168 & \textbf{27.5} & 14.1 \\
5XJY & 6724  & 4.1  & 0.017 & \textbf{0.117} & \textbf{21.1} & 10.0 \\
5FN5 & 3240  & 4.3  & \textbf{0.134} & 0.105 & \textbf{40.3} & 12.3 \\
5WRG & 6679  & 4.3  & 0.081 & \textbf{0.149} & \textbf{10.0} & 7.4  \\
\hline
\end{tabular}
\end{table}

\clearpage 
\end{document}